\magnification=\magstep 1
\overfullrule=0pt
\hfuzz=16pt
\voffset=0.0 true in
\vsize=8.8 true in
\baselineskip 20pt
\parskip 6pt
\hoffset=0.0 true in
\hsize=6.3 true in
\nopagenumbers
\pageno=1
\footline={\hfil -- {\folio} -- \hfil}

\noindent J. Stat. Phys. {\bf 110}, 957-970 (2003)    {\hfill\hfill}   E-print quant-ph/0205037 at www.arxiv.org

\ 

\ 

\centerline{{\bf Initial Decoherence of Open Quantum Systems}}

\vskip 0.16 in

\centerline{{Vladimir Privman}}

\vskip 0.16 in

\centerline{\sl Center for Quantum Device Technology, Clarkson University}
\centerline{\sl Potsdam, New York 13699--5820, USA}

\vskip 0.32 in

\centerline{\bf Abstract}

We present a new short-time approximation scheme for evaluation of decoherence.
At low temperatures, the approximation is argued to apply at intermediate times
as well. It then provides a tractable approach complementary to Markovian-type
approximations, and appropriate for evaluation of deviations from pure
states in quantum computing models.

\vskip 0.32 in


\vskip 0.32 in

\noindent Author's e-mail: privman@clarkson.edu

\vfil\eject

\noindent{\bf 1. Introduction}

Consider a microscopic quantum system with the
Hamiltonian $H_S$. We will refer to the quantum-computing single
quantum bit (qubit) or multi-qubit
paradigm to help define the questions and set up
the challenges, in describing how the system, $S$,
interacts with the surrounding macroscopic world.
However, in principle $S$ can be any quantum system.

Interactions with the surroundings can be
quite different depending on the setting. For example,
in quantum measurement, which for orthodox quantum theory is not fully
understood, the wavefunction of the system is probed,
so part of the process would involve a strong interaction
with the measuring device, such that the system's own
Hamiltonian plays no role in the process. However,
in most applications,
the external interactions are actually
quite weak. Furthermore,
the aim is to minimize their effect, especially in quantum
computing.

Traditionally, interactions with the surrounding world have been
modeled by the modes of a bath, $B$, with each mode described by
its Hamiltonian $M_K$, so that the bath of modes is represented by

$$ H_B = \sum_K M_K \; . \eqno(1.1) $$

\noindent The interaction, $I$, of the bath modes with the system $S$,
will be modeled by

$$ H_I = \Lambda_S P_B = \Lambda_S \sum_K J_K \; , \eqno(1.2) $$

\noindent where $\Lambda_S$ is some Hermitean operator of $S$, coupled to
the operator $P_B$ of the bath.

The bath, or ``heat bath'', can be a collection of modes, such as photons,
phonons, spins, excitons, etc. For a bosonic bath of oscillators, [1-6],
which we use for derivation of specific results, we take

$$ M_K = \omega_K a^\dagger_K a_K \; , \eqno(1.3) $$

$$ J_K = g^*_K a_K + g_K a^\dagger_K \; . \eqno(1.4) $$

\noindent Here we have assumed that the energy of the ground state is
shifted to zero for each oscillator, and we work in
units such that $\hbar =1$.

The total Hamiltonian of the system and bath is

$$ H = H_S + H_B + H_I \; . \eqno(1.5) $$

\noindent More generally, the interaction, (1.2), can involve
several system operators, each coupling differently to the
bath modes, or even to different baths. The bath modes, in turn,
can be coupled to specified external objects, such as impurities.

Let $\rho (t) $ represent the reduced density matrix of the
system at time $t \geq 0$, after the bath modes have been traced over.
For large times, the effect of the environment on a
quantum system that is not otherwise externally controlled,
is expected to be thermalization: the density matrix should
approach

$$ \rho (t \to \infty ) = {\exp \left(-\beta H_S \right)
 \over {\,{\rm Tr}\,}_S \left[\exp \left(-\beta H_S \right) \right] }\; ,
 \eqno(1.6) $$

\noindent where $\beta \equiv 1 / k T $.
At all times, we can consider the degree to
which the system has departed from coherent pure-quantum-state evolution.
This departure is due to the interactions and entanglement with the bath.
We also expect that the temperature, $T$, and other external parameters
that might be needed to characterize the system's density matrix, are
determined by the properties of the bath, which in turn might interact
with the rest of the universe.

Let us introduce the eigenstates of $H_S$,

$$ H_S | n \rangle = E_n | n \rangle \; , \eqno(1.7) $$

\noindent and have $\Delta E$ denote the characteristic energy gap
 values
of $S$. We also consider the matrix elements of $\rho (t)$,

$$ \rho_{mn} (t) = \langle m | \rho (t) | n \rangle \; . \eqno(1.8) $$

\noindent For large times, we expect the diagonal elements $\rho_{nn}$
to approach values proportional to $e^{-\beta E_n}$, while the
off-diagonal elements, $\rho_{m\neq n}$, to vanish. These properties
can be referred to as thermalization and decoherence in the energy basis,
though ``thermalization'' in the strong sense of (1.6) implies decoherence.

To establish these thermalization and decoherence properties,
several assumptions are made regarding the system and bath
dynamics [1-11]. At time $t=0$, it is usually assumed that
the bath modes, $K$, are thermalized, i.e., have density matrices

$$ \theta_K = e^{-\beta M_K} \Big / {\,{\rm Tr}\,}_K \left(
e^{-\beta M_K} \right) \; . \eqno(1.9)$$

\noindent The density matrix $R$ of the system plus bath
at time $t=0$ is assumed to be the direct product

$$  R(0)=\rho(0)\prod_K \theta_K   \; , \eqno(1.10)$$

\noindent and the system and bath modes are not entangled with each other.

Now, a series of assumptions are made, e.g., the Markovian and
secular approximations. The most important is the Markovian approximation,
which, even though it can be stated and introduced in various
ways, essentially assumes that the density matrices of the
bath modes are reset externally to the thermal ones, on
time scales shorter than any dynamical times of the system
interacting with the bath, and the
product form of the full density matrix is maintained. This is a natural assumption,
because each bath mode is coupled only weakly to the
system, whereas it is ``monitored'' by the rest of the
universe and kept at temperature $T$. In its straightforward
version, this amounts to using (1.10) for times $t>0$.
Ultimately, such approaches aim at master equations for
the evolution of $\rho_{mn}(t)$ at large times, consistent
with the Golden Rule and with the expected thermalization
and decoherence properties.

In variants of these formalisms, several time scales are identified.
One is the inverse of the upper cutoff, Debye frequency of the bath
modes, $1/\omega_D$. Another is the thermal time $\hbar / kT = \beta $
(in units of $\hbar =1$). The system $S$ has its own characteristic
time, $1/\Delta E$,
as well as the system-bath dynamical times of thermalization and
decoherence, etc., $T_{1,2,\ldots}$, corresponding to the
``intrinsic'' NMR/ESR times $T_1$, $T_2$, etc.
Heuristically, bath modes of frequencies $\omega$ comparable
to $\Delta E$ are needed to drive thermalization and decoherence.
Initial decoherence can be also mediated by the modes near $\omega = 0$.
At low temperatures, we can assume that $1/\omega_D < 1/\Delta E < \beta$.

There is evidence [7,11,12] that at low temperatures, the
Markovian-type and other approximations used in the derivation
of equations for thermalization and decoherence, are only valid
for times larger than the thermal time scale $\beta$. For quantum
computing applications, in solid-state semiconductor-heterostructure
architectures [13-19], we expect temperatures of several tens of $\mu{}$K.
The thermal time scale then becomes dangerously close to the external
single-qubit control, Rabi-flip time even for slower qubits, those
based on nuclear spins. We emphasize that not all the approximation
schemes have this problem [11].

In Section 2, we offer additional comments on decoherence and
quantum computing. Then, in Section 3, we develop a short-time-decoherence
approximation. In a discussion at the end of Section 3, we offer arguments
that, at low temperatures, our approximation is actually valid for
intermediate times, larger than $1/\omega_D$, hopefully up to times
comparable or larger than $ 1/\Delta E $. Specific results for
the bosonic heat bath are presented in
Section 4. Section 5 comments on the case of adiabatic decoherence,
when the short-time approximation becomes exact.

\vfil\eject

\noindent {\bf 2. Decoherence and quantum computing}

Quantum computing architectures usually emphasize systems, both the qubits
and the modes that couple them (and at the same time act as a bath mediating
unwanted coupling to the rest of the universe), that have large spectral
gaps. It is believed that, especially at low temperatures, spectral gaps
slow down relaxation processes. Therefore, quantum computing architectures
usually assume [13-19] qubits in quantum dots, or in atoms, or subject to
large magnetic fields, and coupled by highly nondissipative quantum media [14,19].

The spectral gaps are expected to slow down exponentially, by the Boltzmann
factor, the processes of thermalization, involving energy exchange. Off-shell
virtual exchanges, will be also slowed down, but less profoundly.
The latter processes contribute to decoherence. Therefore, at low
temperatures, we might expect separation of time scales of the
initial decoherence vs.\ later-stage thermalization and further decoherence.
The latter two processes are described by the traditional NMR/ESR
intrinsic $T_1$ and $T_2$, respectively.

Since only thermalization is clearly associated with the energy eigenbasis,
one can also ask whether the energy basis is the appropriate one
to describe decoherence for short and intermediate times, before
the thermalizing processes, that also further drive decoherence, take over.
The issue of the appropriate basis for studying decoherence, has
also come up in models of quantum measurement. It has been
argued [20-24] that the eigenbasis
of the interaction operator, $\Lambda_S$, may be more
appropriate for intermediate times than the energy eigenbasis.

Yet another aspect of decoherence in quantum computing, involves
the observation that we really want to retain a
{\it pure state\/} in the quantum computation process [25-30].
Decay of off-diagonal matrix elements, in whatever basis, might
not be the best measure of deviations from the pure-state density
matrix, where by pure states we mean those with density matrices
that are projection operators $|\psi\rangle{}\langle\psi|$. 
For instance, the deviation of ${\,{\rm Tr}\,}_S \left[\rho^2 (t)\right]$
from 1, may be more appropriate, and is easer to calculate than
other measures, specifically those motivated by the ``entropic''
expressions proportional to ${\,{\rm Tr}\,}_S \left[\,\rho \ln\left(
\rho\right)\right]$ . Therefore, it is desirable to
have basis-independent expressions for the reduced density operator $\rho (t)$.

Recently, several groups have reported [12,19,24,31-41] results
for spin decoherence in solid state systems appropriate for quantum
computing architectures. Some of these works have not invoked the
full battery of the traditional approximations, Markovian and secular,
etc., or have utilized the spectral gap of the bath modes, to achieve
better reliability of the short-time results. In [41], interaction of
the spin-exciton bath modes with impurities was accounted for, as the
main mechanism of decoherence. In the present work, we limit ourselves
to the bath modes only interacting with the system. Experimental efforts
are picking up momentum, with the first limited results available [42,43]
by traditional NMR/ESR techniques, with the quantum-computing emphasis.

An approach, termed adiabatic decoherence, have been developed by us
[24], expanding the earlier works [12,31-33], with the goal of avoiding
the ambiguity of the basis selection and achieving exact solvability.
The price paid was the assumption that $H_S$ is conserved (a particular
version of the quantum nondemolition processes), which is equivalent to
requiring that

$$ [H_S,H] = [H_S,\Lambda_S] =0 \qquad {\rm (adiabatic\ case)}\; . \eqno(2.1)  $$

\noindent This makes the eigenbasis of $H_S$ and $\Lambda_S$ the same,
but precludes energy relaxation, thus artificially leaving only energy-conserving
relaxation pathways that contribute to decoherence. We will comment on the
results of this approach in Section 5.

Most of the results referred to earlier, have involved approximations
of one sort or another. The most popular and widely used approximation
has been the second-order perturbative expansion in the interaction
strength, $H_I$, though some nonperturbative results have also been
reported. In Section 3, we describe a novel approximation scheme [44]
that is valid for short times. It has several advantages, such as
becoming exact in the adiabatic case, allowing derivation of several
explicit results, and, at least in principle, permitting derivation
of higher-order approximations. Certain models of quantum measurement
evaluate decoherence by effectively setting $H_S=0$. Our approximation
then becomes exact, and our results are consistent with these studies [45,46].

Our formulation in Section 3, will be quite general, and we will not use
the specific bath or thermalization assumptions. However, we do utilize
the factorization property (1.10) at time $t=0$. Thus, we do have to
assume that, at least initially, the system and the bath modes are not
entangled.
In fact, the present formulation also relies on that the Hamiltonians
at hand are all time-independent. Therefore, we have excluded the
possibility of controlled dynamics, in the quantum computing sense,
when gate functions are accomplished by external couplings to
individual qubits and by external control of their pairwise
interactions. Our formulation, therefore, applies to ``idling''
qubits or systems of (possibly interacting) qubits. It is reasonable
to assume that a lower limit on decoherence
rate can be evaluated in such an idling state, even though for
quantum error correction, qubits otherwise idling, might be
frequently probed (measured) and entangled with ancillary qubits [25-30].

The $t=0$ factorization assumption (1.10), shared by all the
recent spin-decoherence studies, then represents the expectation
that external control by short-duration but large externally applied
potentials, measurement, etc., will ``reset'' the qubits, disentangling
them from the environment modes to which the affected qubits are only
weakly coupled. Thus, we assert that it is the qubit system that gets
approximately reset and disentangled
from the bath towards time $t=0$, instead of the bath being thermalized by the
rest of the universe, as assumed in Markovian approximation schemes.

\vfil\eject

\noindent {\bf 3. Short-time decoherence}

In addition to the energy basis, (1.7), we also define the eigenstates of the
interaction operator $\Lambda_S$, by

$$ \Lambda_S | \gamma \rangle = \lambda_\gamma | \gamma \rangle \; , \eqno(3.1) $$

\noindent where the Greek index labels the eigenstates of $\Lambda_S$, with
eigenvalues $\lambda_\gamma$, while the Roman indices will be used for the
energy basis, and, when capitalized, for the bath modes, (1.2)-(1.4).

The time dependence of the density matrix $R(t)$ of the system and bath, is formally given by

$$ R(t) = e^{-i(H_S+H_B+H_I)t} R(0) \, e^{i(H_S+H_B+H_I)t} \; . \eqno(3.2) $$

\noindent We will utilize the following approximate relation for the exponential
factors, as our short-time approximation,

$$ e^{i(H_S+H_B+H_I)t + O(t^3)} = e^{iH_St/2}\, e^{i(H_B+H_I)t}\, e^{iH_St/2} \; . \eqno(3.3)$$

\noindent This relation has the following appealing properties. It becomes
exact for the adiabatic case, (2.1). Furthermore, if we use the right-hand
side and its inverse to replace $e^{\pm iHt}$, then we are imposing three
time-evolution-type transformations on $R(0)$. Therefore, the
approximate expression for $R(t)$ will have all the desired properties
of a density operator. Finally, extensions to higher-order approximations
in powers of $t$ are possible, by using relations derived
in [47], where various expressions valid to $O(t^4)$ and $O(t^5)$ were considered.

Our goal is to evaluate the resulting approximation to the matrix element,

$$ \rho_{mn}(t) = {\,{\rm Tr}\,}_B \langle m | e^{-iH_St/2}\, e^{-i(H_B+H_I)t}\,
e^{-iH_St/2} R(0) \, e^{iH_St/2}\, e^{i(H_B+H_I)t}\, e^{iH_St/2} |n
\rangle \; . \eqno(3.4) $$

\noindent First, we apply the operators
$H_S$ in the outer exponentials, acting to the left on $\langle m |$ ,
and to the right on $|n
\rangle $, replacing $H_S$ by, respectively,
$E_m$ and $E_n$.
We then note that the second exponential operator in (3.4) contains
$\Lambda_S $, see (1.2). Therefore, we insert the decomposition
of the unit operator in the system space, in terms of the eigenbasis of $\Lambda_S$, before
the second exponential, and one in terms of the eigenbasis of $H_S$
after it. This allows us to apply $\Lambda_S$ in the second exponential and also
$H_S$ in the third exponential. The same substitution is carried out
on the other side of $R(0)$, with the result

$$ \rho_{mn}(t) = \sum_{\gamma\, p\, q\, \delta}
{\,{\rm Tr}\,}_B \Big[\,e^{-iE_m t/2}\langle m |\gamma \rangle
\langle \gamma |p \rangle e^{-i(H_B+\lambda_\gamma P_B)t}\,
e^{-iE_p t/2} \rho_{pq}(0) $$
$$\times\Big( \prod_K \theta_K \Big)
e^{iE_q t/2} \, e^{i(H_B+\lambda_\delta P_B)t}
\langle q |\delta \rangle \langle \delta |n \rangle
e^{iE_n t/2}\,\Big] \; . \eqno(3.5) $$

The next step is to collect all the terms, and also identify that the
trace over the bath can be now carried out for each mode separately. We use (1.1)-(1.2) to write

$$ \rho_{mn}(t) = \sum_{\gamma\, p\, q\, \delta}\Big\{
e^{i(E_q+E_n-E_p-E_m)t/2}\langle m |\gamma \rangle
\langle \gamma |p    \rangle \, \rho_{pq}(0)\,\langle q |\delta \rangle \langle \delta |n \rangle  $$
$$\times
\prod_K {\,{\rm Tr}\,}_K  \Big[ e^{-i(M_K+\lambda_\gamma J_K)t}\,\theta_K\,
e^{i(M_K+\lambda_\delta J_K)t} \Big] \Big\} \; . \eqno(3.6) $$

\noindent While this expression looks formidable, it actually
allows rather straightforward calculations in some cases. Specifically,
the simplest quantum-computing applications involve two-state systems. Then the sums
in (3.6) are over two terms each. The calculations involving the
overlap Dirac brackets between the eigenstates of $H_S$ (labeled by $m$,
$n$, $p$ and $q$) and of $\Lambda_S$ (labeled by $\gamma$ and $\delta$),
as well as the energy-basis matrix elements of $\rho(0)$, cf.\ (1.8),
involve at most diagonalization of two-by-two Hermitean matrices. Of
course, the approximation (3.6) can be used for evaluation of short-time
density matrices for systems more general than two-state.

The challenging part of the calculation involves
the trace over each mode of the bath. Since these modes have identical
structure, e.g., (1.3)-(1.4) for the bosonic bath case, but with $K$-dependent
coupling constants, the calculation needs only be done once, in the space of
{\it one mode}. Furthermore, results for the bath models ordinarily used, such
as the bosonic and spin baths, are either already available in the literature
or can be calculated without much difficulty. For the  thermalized initial
bath-mode density matrix $\theta_K$, we give the exact bosonic-model expression in the next section.

In the remainder of this section, we first further analyze the trace over
one bath mode entering (3.6). We then comment on the limits of validity of the present approximation.

In an obvious shorthand notation, we write the single-mode trace in (3.6) as

$$ {\,{\rm Tr}\,}  \Big[ e^{-i(M+\gamma J)t}\,\theta\,
e^{i(M+\delta J)t} \Big]  =
{\,{\rm Tr}\,}  \Big[ \,\theta\,
e^{i(M+\delta J)t} \,e^{-i(M+\gamma J)t}\Big] \; . \eqno(3.7) $$

\noindent Now, to the same order of approximation as used in (3.3), we
can write

$$ e^{i(M+\delta J)t + O(t^3)} = e^{iMt/2}\, e^{i\delta Jt}\, e^{iMt/2} \; . \eqno(3.8)$$

\noindent The resulting approximation for the trace (3.7) reads

$$ {\,{\rm Tr}\,}  \Big[ \Big (e^{-iMt/2}\,\theta\,
e^{iMt/2}\big) e^{i(\delta - \gamma)Jt} \Big] \; , \eqno(3.9) $$

\noindent which illustrates that, within this approximation, the product of traces in
(3.6) is a function of the difference $\lambda_\gamma - \lambda_\delta $.
In fact, this product is exactly 1 for $\lambda_\gamma = \lambda_\delta $ and,
in most applications, the following form is likely to emerge,

$$ \prod_K {\,{\rm Tr}\,}_K [\ldots] = e^{-\,{\rm const}\, \left(\lambda_\gamma -
\lambda_\delta\right)^2t^2 + O\left(t^3\right)} \; , \eqno(3.10) $$

\noindent though we caution the reader that (3.10) is somewhat speculative
and suggested by the exact result for the bosonic heat bath, reported in
the next section.

Finally, we point out that in most cases of interest, the initial single-mode
density matrix $\theta$ will commute with the bath-mode energy operator $M$.
In fact, the thermalized $\theta$ is a function of $M$.  Therefore,
(3.9) can be further simplified to

$$ {\,{\rm Tr}\,}  \Big[ \,\theta\,
e^{i(\delta - \gamma)Jt} \Big] \; . \eqno(3.11) $$

However, let us emphasize that the approximate relations (3.9)-(3.11) are
likely of value only as far as they help to derive basis-independent (operator)
approximations to $\rho (t)$, by a technique illustrated in the next section.
Indeed, for most bath models it is advisable to calculate the single-mode
trace exactly first, according to
(3.6), and then attempt various approximations.

The latter statement reflects our expectation that the approximation
developed here is valid, for low temperatures, not only for short
times, defined by $t < 1/ \omega_D$, but also for intermediate
times, exceeding $ 1/\omega_D $. This is suggested
by the result of an illustrative calculation in the next
section, but mainly by the fact that (3.11) only includes
the bath-mode energy scales via $\theta$, and, therefore,
at low temperatures, is dominated by the lowest bath-mode
excitations, and is not sensitive to frequencies of order
$\omega_D$. Thus, we expect our approximation to be applicable
complementary to the Markovian-type approximations and definitely
break down in the regime of fully developed thermalization, for $t
\geq O(\beta)$. Additional supporting observations are offered in Section
5, when we consider the adiabatic case (2.1).

\vfil\eject

\noindent {\bf 4. The bosonic heat bath}

In this section, we consider the bosonic heat bath [6], see (1.3)-(1.4),
in the initially thermalized state,

$$\theta_K = e^{-\beta M_K} / {\,{\rm Tr}\,}_K \left ( e^{-\beta M_K} \right ) =
\left( 1 - e^{-\beta \omega_K} \right ) e^{-\beta \omega_K a^\dagger_K a_K} \; . \eqno(4.1) $$

\noindent The product of the single-mode traces in (3.6), is then available in the
literature [12,24,31],

$$ \rho_{mn}(t) = \sum_{\gamma\, p\, q\, \delta}\Big\{
e^{i(E_q+E_n-E_p-E_m)t/2}\langle m |\gamma \rangle
\langle \gamma |p \rangle \langle q |\delta \rangle \langle \delta |n \rangle \rho_{pq}(0) $$
$$\times
\exp \Big( - \sum_K { |g_K|^2 \over \omega_K^2 } \Big [
2 \left(\lambda_\gamma - \lambda_\delta \right)^2
\sin^2 {\omega_K t \over 2} \coth {\beta \omega_K \over 2}
 +\, i \left(\lambda_\gamma^2 - \lambda_\delta^2 \right) \left(
\sin \omega_K t - \omega_K t \right) \Big ] \Big) \Big \} \; . \eqno(4.2) $$

\noindent The last term in the exponent, linear in $t$, is usually viewed as
``renormalization'' of the system energy levels due to its interaction
with the bath modes. It can be removed by adding the term,

$$H_R= \Lambda_S^2 \sum_K |g_K|^2 / \omega_K \; , \eqno(4.3) $$

\noindent to the total Hamiltonian. However, the usefulness of this
identification for short times
is not clear, and we will not use it. One can check that, {\it unmodified},
(4.2) is consistent with the expectation (3.10).

Let us now define two non-negative real spectral sums, $B(t)$ and $C(t)$, over the bath modes,

$$ B^2(t) = 8 \sum_K { |g_K|^2 \over \omega_K^2 }
\sin^2 {\omega_K t \over 2} \coth {\beta \omega_K \over 2}
 \; , \eqno(4.4) $$

$$ C(t) = \sum_K { |g_K|^2 \over \omega_K^2 } \left (
\omega_K t - \sin \omega_K t   \right) \; . \eqno(4.5) $$

\noindent When converted to integrals
over the bath mode frequencies, with the cutoff
at $\omega_D$, these sums have been discussed extensively in the
literature [6,12,31], for several choices of the bath mode density
of states and coupling strength $g$ as functions of
the mode frequency.

The final expression is,

$$ \rho_{mn}(t) = \sum_{\gamma\, p\, q\, \delta}\Bigg\{
e^{i(E_q+E_n-E_p-E_m)t/2}\langle m |\gamma \rangle
\langle \gamma |p \rangle \langle q |\delta \rangle \langle \delta |n \rangle \rho_{pq}(0) $$
$$\times
\exp \left[ - {1\over 4}B^2(t) \left(\lambda_\gamma - \lambda_\delta \right)^2
\,+\, i C(t) \left(\lambda_\gamma^2 - \lambda_\delta^2 \right) \right] \Bigg \} \; . \eqno(4.6) $$

\noindent When the spectral functions are expanded in powers of $t$, this
result confirms all the conclusions and conjectures discussed in Section
3, in connection with relations (3.9)-(3.11).

Let us now turn to the derivation of the basis-independent representation
for $\rho (t)$, by utilizing the integral identity

$$ \sqrt{\pi} \exp [ - B^2 (\Delta\lambda)^2 /4] = \int_{-\infty}^\infty \!
dy \, e^{-y^2} \exp [ i y B (\Delta\lambda)] \; . \eqno(4.7) $$

\noindent Exponential factors in (4.6) can then be reproduced by applying
operators on the wavefunctions entering the overlap Dirac brackets, with
the result

$$ \sqrt{\pi} \,\rho (t)=  \int\!
dy \, e^{-y^2} e^{-iH_St/2} \,e^{i[yB(t)\Lambda_S+C(t)\Lambda_S^2]}
\,e^{-iH_St/2} \,\rho(0) \,e^{iH_St/2} \,e^{-i[yB(t)\Lambda_S+C(t)\Lambda_S^2]} \,e^{iH_St/2} \; . \eqno(4.8) $$

Within the $O(t^2)$ approximation (3.3), given that $B$ and $C$ are of order
linear or higher in $t$, we can combine the exponential operators to get an alternative approximation,

$$ \sqrt{\pi} \,\rho (t) =  \int\!
dy \, e^{-y^2} \,e^{-i[tH_S-yB(t)\Lambda_S-C(t)\Lambda_S^2]}
 \,\rho(0) \,e^{i[tH_S-yB(t)\Lambda_S-C(t)\Lambda_S^2]} \; , \eqno(4.9) $$

\noindent though (4.6) and (4.8) are in fact easier to handle in actual calculations.

As an application, let us consider the case of $H_S$ proportional to the
Pauli matrix $\sigma_z$, e.g., a spin-$1/2$ particle in magnetic field,
and $\Lambda_S = \sigma_x$, with the proportionality constant in the
latter relation absorbed in the definition of the coupling constants
$g_K$ in (1.4). Let us study the deviation of the state of a spin-$1/2$
qubit, initially in the energy eigenstate $|\uparrow\,\rangle$ or
$|\downarrow\,\rangle$, from pure state, by calculating ${\,{\rm Tr}\,}_S \, [\rho^2 (t)]$
according to (4.8). We note that
for a two-by-two density matrix, this trace can vary from 1
for pure quantum states to the lowest value of $1/2$ for maximally mixed states.

A straightforward calculation with $\rho (0) = |\uparrow\,\rangle \langle
\,\uparrow |$ or $|\downarrow\,\rangle \langle \,\downarrow |$, yields

$$ {\,{\rm Tr}\,}_S \, [\rho^2 (t)] = {1\over 2} \left[1+e^{-2B^2(t)}\right] \; . \eqno(4.10)$$

\noindent As the time increases, the function $B^2(t)$ grows
monotonically from zero [6,12,24,31]. Specifically, for Ohmic dissipation,
$B^2(t)$ increases quadratically for short times $t < O(1/\omega_D)$, then
logarithmically for $O(1/\omega_D) < t < O(\hbar /kT)$, and linearly for
$t>O(\hbar/kT)$. For other bath models, it need not diverge to infinity
at large times.

Both approximations, (4.8)-(4.9), make the deviation from a pure state
$\rho(0)=|\psi_0\rangle\langle\psi_0|$ apparent: $\rho (t>0)$ is obviously
a {\it mixture\/} (integral over $y$) of pure-state projectors
$|\psi(y,t)\rangle\langle\psi(y,t)|$, where, for instance for (4.9),

$$\psi(y,t)= e^{-i[tH_S-yB(t)\Lambda_S-C(t)\Lambda_S^2]} \,\psi_0 \; , \eqno(4.11) $$

\noindent with a somewhat different expression for (4.8).

\vfil\eject

\noindent {\bf 5. The adiabatic case}

Relation (2.1) corresponds to the system's energy conservation. Therefore,
energy flow in and out of the system is not possible, and normal
thermalization mechanisms are blocked. This ``adiabatic decoherence'' limit
thus corresponds to ``pure dephasing''; see [48].

The fact that our approximation
becomes exact in this case, provides support to the expectation that,
at low temperatures, it is generally valid beyond the cutoff time scale
$1/\omega_D$, providing a reasonable evaluation of decoherence and
deviation from a pure state, as exemplified by the calculation yielding
(4.10), in Section 4.

With (2.1), we can select a common eigenbasis for $H_S$ and $\Lambda_S$.
Then the distinction between the lower-case Roman and Greek indices in (3.6) becomes
irrelevant, and the sums can all be evaluated to yield

$$ \rho_{mn}(t) =
e^{i(E_n-E_m)t}\, \rho_{mn}(0)\,
\prod_K {\,{\rm Tr}\,}_K  \Big[ e^{-i(M_K+\lambda_m J_K)t}\,\theta_K\,
e^{i(M_K+\lambda_n J_K)t} \Big]  \; . \eqno(5.1) $$

\noindent This expression was discussed in detail in our work on
adiabatic decoherence [24]. Specifically, for the initially
thermalized bosonic heat bath case, we have, for the absolute
values of the density matrix elements,

$$ \big | \rho_{mn}(t) \big | = \big | \rho_{mn}(0) \big |\, e^{  - B^2(t)
(\lambda_m - \lambda_n )^2 /4 } \; . \eqno(5.2) $$

\noindent The decay of the off-diagonal matrix elements thus depends of the
properties of the spectral function $B^2(t)$ as the time increases.
Such explicit results [12,24,31-33] illustrate that for 
irreversibile behavior, the number of bath modes must be infinite,
with the spectral function evaluated in the continuum limit.

In summary, we have derived short-time approximations for
the density matrix and its energy-basis matrix elements. Our
expressions are quite easy to work with, because for few-qubit
systems they only involve manipulation of finite-dimensional matrices,
and they will be useful in estimating decoherence and deviation from
pure states in quantum computing models, including results for low temperatures.

This research was supported by the National Science Foundation, grants
DMR-0121146 and ECS-0102500, and by the National Security Agency and
Advanced Research and Development Activity under Army Research Office
contract DAAD-19-99-1-0342.

\vfil\eject

\centerline{\bf References}{\frenchspacing

\item{1.} R.P. Feynman and A.R. Hibbs,
{\it Quantum Mechanics and Path Integrals\/}
(McGraw-Hill, NY, 1965).

\item{2.} G.W. Ford, M. Kac and P. Mazur,
J. Math. Phys. {\bf 6}, 504 (1965).

\item{3.} A.O. Caldeira and A.J. Leggett,
Phys. Rev. Lett. {\bf 46}, 211 (1981).

\item{4.} A.O. Caldeira and A.J. Leggett, Physica {\bf 121A},
587 (1983)

\item{5.} S. Chakravarty and A.J. Leggett, Phys. Rev. Lett.
{\bf 52}, 5 (1984).

\item{6.} A.J. Leggett, S. Chakravarty, A.T. Dorsey,
M.P.A. Fisher and W. Zwerger, Rev. Mod. Phys. {\bf 59}, 1
(1987) [Erratum {\it ibid.\/} {\bf 67}, 725 (1995)].

\item{7.} N.G. van Kampen, {\it Stochastic Processes in Physics and
Chemistry\/} (North-Holland, Amsterdam, 2001).

\item{8.} W.H. Louisell, {\it Quantum Statistical
Properties of Radiation\/} (Wiley, NY, 1973).

\item{9.} K. Blum, {\it Density Matrix Theory and
Applications\/} (Plenum, NY, 1996).

\item{10.} A. Abragam, {\it The Principles of Nuclear Magnetism\/} (Clarendon Press, 1983).

\item{11.} H. Grabert, P. Schramm and G.-L. Ingold,
Phys. Rep. {\bf 168}, 115 (1988).

\item{12.} N.G. van Kampen, J. Stat. Phys. {\bf 78},
299 (1995).

\item{13.} D. Loss and D.P. DiVincenzo, Phys. Rev. {\bf A57}, 120 (1998).

\item{14.}  V. Privman, I.D. Vagner and G. Kventsel, Phys. Lett. {\bf A239}, 141 (1998).

\item{15.} B.E. Kane, Nature {\bf 393}, 133 (1998).

\item{16.} A. Imamoglu, D.D. Awschalom, G. Burkard, D.P. DiVincenzo,
D. Loss, M. Sherwin and A. Small, Phys. Rev. Lett. {\bf 83}, 4204 (1999).

\item{17.} R. Vrijen, E. Yablonovitch, K. Wang, H.W. Jiang, A. Balandin,
V. Roychowdhury, T. Mor and D.P. DiVincenzo, Phys. Rev. {\bf A62}, 012306 (2000)

\item{18.} S. Bandyopadhyay, Phys. Rev. {\bf B61}, 13813 (2000).

\item{19.} D. Mozyrsky, V. Privman and M.L. Glasser, Phys. Rev. Lett. {\bf 86}, 5112 (2001).

\item{20.} W.G. Unruh and W.H. Zurek, Phys. Rev. {\bf D40},
1071 (1989).

\item{21.} W.H. Zurek, S. Habib and J.P. Paz,
Phys. Rev. Lett. {\bf 70}, 1187 (1993).

\item{22.} A.O. Caldeira and A.J. Leggett,
Ann. Phys. {\bf 149}, 374 (1983).

\item{23.} L. Mandel and E. Wolf, {\it Optical Coherence
and Quantum Optics\/} (Cambridge University Press, 1995).

\item{24.} D. Mozyrsky and V. Privman, J. Stat. Phys. {\bf 91},
787 (1998).

\item{25.} P.W. Shor, in {\it Proc. 37th Annual Symp. Found. Comp. Sci.},
p. 56 (IEEE Comp. Sci. Soc. Press, Los Alamitos, CA, 1996).

\item{26.} D. Aharonov and M. Ben-Or, {\it Fault-Tolerant Quantum
Computation with Constant Error}, preprints quant-ph/9611025 and quant-ph/9906129.

\item{27.} A. Steane, Phys. Rev. Lett. {\bf 78}, 2252 (1997).

\item{28.} E. Knill and R. Laflamme, Phys. Rev. {\bf A55}, 900 (1997).

\item{29.} D. Gottesman, Phys. Rev. {\bf A57}, 127 (1998).

\item{30.} J. Preskill, Proc. Royal Soc. London {\bf A454}, 385 (1998).

\item{31.} G.M. Palma, K.A. Suominen and A.K. Ekert,
Proc. Royal Soc. London {\bf A452}, 567 (1996).

\item{32.} J. Shao, M.-L. Ge and H. Cheng, Phys. Rev. {\bf E53},
1243 (1996).

\item{33.} I.S. Tupitsyn, N.V. Prokof'ev, P.C.E. Stamp,
Int. J. Mod. Phys. {\bf B11}, 2901 (1997).

\item{34.} N.V. Prokof'ev and P.C.E. Stamp,
Rep. Prog. Phys. {\bf 63}, 669 (2000).

\item{35.} J. Ankerhold and H. Grabert, Phys. Rev. {\bf E61},
3450 (2000).

\item{36.} T. Maniv, Y.A. Bychkov, I.D. Vagner and P. Wyder, Phys. Rev. {\bf B64}, 193306 (2001).

\item{37.} A.M. Dyugaev, I.D. Vagner and P. Wyder, {\it On the
Electron Scattering and Dephasing by the Nuclear Spins}, preprint cond-mat/0005005.

\item{38.} D. Mozyrsky, S. Kogan and G.P. Berman, {\it Time Scales
of Phonon Induced Decoherence of Semiconductor Spin Qubits}, preprint cond-mat/0112135.

\item{39.} A.V. Khaetskii, D. Loss and L. Glazman, {\it Electron
Spin Decoherence in Quantum Dots due to Interaction with Nuclei}, preprint
cond-mat/0201303.

\item{40.} I.A. Merkulov, A.L. Efros and M. Rosen, {\it Electron
Spin Relaxation by Nuclei in Semiconductor Quantum Dots}, preprint cond-mat/0202271.

\item{41.} D. Mozyrsky, V. Privman and I.D. Vagner, Phys. Rev. {\bf B63}, 085313 (2001).

\item{42.} J. Zhang, Z. Lu, L. Shan and Z. Deng, {\it Experimental Study of
Quantum Decoherence using Nuclear Magnetic Resonance}, preprint quant-ph/0202146.

\item{43.} E. Yablonovitch, private communication.

\item{44.} V. Privman, {\it Short-Time Decoherence and Deviation from
Pure Quantum States}, preprint cond-mat/0203039.

\item{45.} D. Mozyrsky and V. Privman, Mod. Phys. Lett. {\bf B14}, 303 (2000).

\item{46.} D. Braun, F. Haake, W.T. Strunz, Phys. Rev. Lett. {\bf 86} 2913 (2001).

\item{47.} A.T. Sornborger and E.D. Stewart, Phys. Rev. {\bf A60},
1956 (1999).}

\item{48.} J.L. Skinner and D. Hsu, J. Phys. Chem. {\bf 90}, 4931 (1986).

\bye